\documentstyle[11pt]{article}
 
\def\pa{\partial}

\def\g{\gamma} 
\def\a{\alpha}
\def\b{\beta}

\def\l{\lambda} 
\def\m{\mu}
\def\n{\nu}

\def\mn{{\mu\nu}}
\def\be{\begin{equation}}
\def\ee{\end{equation}}
\begin{document}

\begin{flushright}
BRX TH-517
\end{flushright}

\begin{center}

{\large\bf A Note on Matter Superenergy Tensors}

S. Deser\\
Brandeis University \\
Department of Physics \\
Waltham, MA 02454, USA\\
E-mail: deser@brandeis.edu

\end{center}

%\renewcommand{\baselinestretch}{3}
%\small \normalsize

%\maketitle

\begin{abstract}We consider Bel--Robinson-like higher derivative
conserved two-index tensors $H_\mn$ in simple matter models,
following a recently suggested Maxwell field version.  In flat
space, we show that they are essentially equivalent to the true
stress-tensors.  In curved Ricci-flat backgrounds it is possible
to redefine $H_\mn$ so as to overcome non-commutativity of
covariant derivatives, and maintain conservation, but they become
model- and dimension- dependent, and generally lose their simple
``BR" form.
\end{abstract}

Historically, the nonexistence of a local stress tensor in
generally covariant theories such as Einstein's led to a
successful search for the next-best thing, the covariantly
conserved but four-index and higher derivative Bel--Robinson (BR)
tensor \cite{002}, quadratic in curvature.  Being traceless in
D=4, it has no ``$T_\mn$-like" 2-index contraction. However, this
discovery led ineluctably to an equally successful search
\cite{003} for matter analogs of BR, despite the presence of
perfectly good $T_\mn$ there.  These quantities resemble BR in
being of higher derivative order and quadratic in the
``curvatures" of the corresponding fields.  In particular, it has
recently been shown \cite{004} that there is a natural 2-index
conserved BR-version of the Maxwell tensor. In flat space QFT,
operators $H_\mn$ whose matrix elements behave like those of the
stress-tensor are essentially proportional to it \cite{001}:  In
momentum space, one expects a conserved symmetric 2-tensor to have
the form $f(q^2) T_\mn (q)$, up to (trivial) identically conserved
terms. By continuity, one might expect some similar property to
hold, at least for test fields, {\it i.e.}, in Ricci-flat spaces.
The results we obtain here confirm these expectations, at least
for $H_\mn$ of simple free field models. Using the simplest --
scalar and vector -- free field models we will first investigate
their $H_\mn$ in flat space and immediately verify the expectation
that they (and their obvious generalizations) are indeed related
to the corresponding $T_\mn$ by form factors.  In gravitational
backgrounds we find that while $H_\mn$ can be redefined to survive
non-commutation of derivatives at least in Ricci-flat spaces,
generically they lose their flat space attributes and become
model- and dimension-dependent.

In flat space, the new tensors are respectively
\be%1
H^s_\mn \equiv \phi_{\m\a} \phi_\n~^\a - \textstyle{\frac{1}{2}}\;
\eta_\mn (\phi_{\a\b}\phi^{\a\b}) \; , \;\;\; \phi_{\m\a} \equiv
\pa^2_{\a\m} \phi \ee
 and \cite{003}
\be%2
H^v_\mn \equiv F_{\m\l\a} F_\n~^{\l\a} - \textstyle{\frac{1}{4}}\;
\eta_\mn (F_{\l\b\a}F^{\l\b\a}) \; , \;\;\; F_{\m\l\a} \equiv
\pa_\a \; F_{\m\l} \ee
 for scalar and Maxwell fields, and exhibit the BR form, in
terms of the ``curvatures" $\phi_\mn$ and $F_{\m\l\a}$. [The extra
derivatives on the fields are clearly completely transparent to
taking divergences, symmetry, etc.]  Our main point, however, is
that, equally obviously, they are simply related to their
corresponding stress tensors, through
\be%3
H^s_\mn = \textstyle{\frac{1}{2}} \; \Box T^s_\mn \ee and
\be%4
H^v_\mn = \textstyle{\frac{1}{2}} \; \Box T^{\rm max}_\mn \; . \ee
Here and throughout we work entirely on-shell, since that is all
that matters; in particular, we used the Maxwell field equation
 \be %5
 \Box F_\mn = 0 \; .
 \ee
 These examples, then, show that the $H_\mn$ are precisely of the
(momentum space) form $f (q^2) T_\mn (q)$, devoid of independent
content.

An excursion into curved space is less directly motivated, as
there is no expected simple physical ``$H$--$T$" connection there.
Still, it is mildly interesting that any results at all can be
drawn in more general backgrounds.  It is immediately clear that
not only do the definitions (1,2) of $H$ not imply (3,4) but that
in neither form does (covariant) conservation of $H$ follow from
that of $T$. The reason is of course the non-commutativity among
covariant derivatives, particularly $[D_\m , \Box] \neq 0$.

A resolution of this impasse exists, however, for spaces that are
Ricci-flat: when $\Box$ is replaced by the Lichnerowicz operator
\cite{005},
\be%6
\tilde{H}_\mn \equiv (LT)_{\mn} \equiv \Box T_\mn + 2R_{\m\a\n\b}
T^{\a\b} \; , \ee
 commutativity is restored:
 \be%7
D^\m (LT)_\mn = L (D^\m T_\mn ) = 0 \; . \ee
 The curvature addition in (6) compensates for $[D^\m , \Box ]
\neq 0$.  This happens in general for and only for Ricci-flat
spaces, as is easily verified (recall that the divergence of the
Riemann tensor is the curl of Ricci). We have thus rescued, albeit
in restricted geometries,\footnote{In D=4, $H^{\rm max}_\mn$ is
also conserved \cite{004} on combined Einstein--Maxwell shell,
{\it i.e.}, for spaces whose Ricci tensor is proportional to
$T^{\rm max}_\mn$, due to the special properties of the Maxwell
tensor in D=4.  This does not hold in other D, nor  even in D=4,
for the scalar model.} a conserved  -- redefined --
$\tilde{H}_\mn$, but at the price of losing the desired BR form
(1,2).  Can the (necessary) shift from $\Box$ to $L$ be made
compatible with (1,2)?  The answer is a very qualified ``yes", for
D=4 Maxwell only \cite{004} and not otherwise.  First we see why
it does not work for scalars:  the only difference between (3) and
(1) is that the former contains extra terms $\sim \Box D_\m \phi$
(of course all derivative ordering must be carefully kept here!).
But $[\Box , D_\m ] \phi \sim D_\m \Box \phi + R_{\m\a} \phi_\a$
and involves only the Ricci tensor, so we can drop it. This
equality of (3) and (1), however, is also its drawback: since only
$\tilde{H} = \Box T + RT$ is conserved, it is not (1) alone, but
the sum of (1) with $RT$ that is conserved. The Maxwell model,
though only in D=4, circumvents the above difficulty precisely
because (4) differs from (2) by terms of the form $F(\Box F)$ and
these no longer vanish.  Instead,
\be%8
 0 = D^\a [D_\a F_\mn + D_\n F_{\a\m} + D_\m F_{\m\a}] = \Box
 F_\mn + \left\{ R^\a ~_{\n\l\m} F_\a\!\!~^\l - (\n\m )\right\} \; .
 \ee
 The ensuing $RFF$ terms in reducing (4) to (2), together with the
 $RT^{\rm max}$ part of $LT^{\rm max}$, cancel each other owing to
 a specifically D=4 Weyl tensor identity \cite{004}; the
 (covariantized) $H^v_\mn$ of (2), being the reduction of
 $\tilde{H}^v_\mn$, is conserved.

Finally, a remark on the original, gravitational, BR tensor.  In
the present context, one might hope to construct it in terms of an
underlying gauge-variant stress-tensor.  This is impossible for a
seemingly accidental reason: the trace of $B_{\mn\l\b}$ vanishes
in D=4 and so there is no $H_\mn$-like candidate. [The only
quadratic on-shell 2-index tensor is the  square of Weyl,
$C_{\m\a\b\g}C^{\n\a\b\g}$, which is a pure trace  in D=4.]  More
generally, this impossibility is understandable  from the known
fact \cite{006} that (already in flat space) all  $T_\mn$
candidates for free linear spin $>$1 gauge fields vary by a
superpotential term, under gauge transformations, one that cannot
be removed by any curls or $\Box$ operators: Hence $\Box T_\mn$
(or more general operators on $T_\mn$) is still conserved but
remains gauge-variant, even on shell.  The required BR process
here is more radical, taking individual ``curls" of the factors in
$T$, to reach the required $B \sim RR$ form.

In summary, we have seen that in flat space, ostensibly novel
conserved symmetric BR-like extensions of matter $T_\mn$ are
really equivalent to it.  In curved but Ricci-flat backgrounds,
conservation of at least the extended $\tilde{H}_\mn = (LT)_\mn$,
can be preserved.  Even then (for all but the Maxwell field in
D=4), since $(LT)$ is no longer of pure BR form (1) or (2), but
acquires various non-minimal additions.

I acknowledge correspondence with the authors of \cite{004}. This
work was supported by NSF grant PHY99-73935.


\begin{thebibliography}{999}
\bibitem{002}
L.\ Bel, C.R.\ Acad.\ Sci., Paris {\bf 248}, 1292 (1959); I.\
Robinson, unpublished lectures, Kings College, London
(1958).
\bibitem{003} M.\ Chevreton, Nuovo Cimento {\bf 34}, 90
(1964).
\bibitem{004} G.\ Bergqvist, I.\ Eriksson and J.M.M.\
Senovilla, gr-qc/0303046.
\bibitem{001} See for example C.A.\ Orzalesi, J.\ Sucher
and C.H.\ Woo, Phys.\ Rev.\ Letters {\bf 21}, 1550 (1968); D.R.\
Divgi and C.H.\ Woo, Phys.\ Rev.\ Letters {\bf 23}, 1510 (1969).
\bibitem{005}
A.\ Lichnerowicz, Publ.\ Math.\ IHES {\bf 10}, 293 (1961).
\bibitem{006}
S.\ Deser and J.\ McCarthy, Class. Quant. Grav. {\bf 7}, L119
(1990).
\end{thebibliography}
\end{document}